\documentclass{ws-procs975x65}
\def\lsim{\lower.5ex\hbox{$\; \buildrel < \over \sim \;$}}
\def\gsim{\lower.5ex\hbox{$\; \buildrel > \over \sim \;$}}
\def\be{\begin{equation}}
\def\ee{\end{equation}}

\begin{document}

\title{QPOs from Radial and Vertical Oscillation of Shocks in Advective Accretion Flows}

\author{SANDIP K. CHAKRABARTI\footnote{\uppercase{A}lso at \uppercase{C}entre for 
\uppercase{S}pace \uppercase{P}hysics, \uppercase{C}halantika 43, \uppercase{G}aria 
\uppercase{S}tation \uppercase{R}d., \uppercase{K}olkata 700084}}
\address{S.N. Bose National Centre for Basic Sciences,\\
JD-Block, Salt Lake, Kolkata 700098\\
E-mail: chakraba@bose.res.in}

\author{KINSUK ACHARYYA}
\address{Centre for Space Physics, Chalantika 43, Garia Station Rd., 
Garia, Kolkata, 700084\\ e-mail:space\_phys@vsnl.com}

\author{DIEGO MOLTENI}
\address{Dipartimento di Fisica e Tecnologie Relative,\\
Viale delle Scienze, 90128 Palermo, Italy; e-mail:molteni@unipa.it}

\maketitle

\abstracts {We present results of several numerical simulations of two dimensional
advective flows which include cooling processes. We show that the computed light curve 
is similar to the $\chi$ state in GRS 1915+105. The power density spectrum (PDS)
also shows presence of QPOs near the break frequency.}

\noindent To be Published in the proceedings of the 10th Marcel Grossman Meeting (World Scientific Co., Singapore),
Ed. R. Ruffini et al.

\section{Introduction}

It has become abundantly clear that Centrifugal pressure dominated BOundary Layers or so-called
CENBOLs, which are formed in between the black hole horizon and the accretion shock are
the most important ingredients in black hole astrophysics (Chakrabarti, this volume). Chakrabarti
and Titarchuk (1995) showed that this is nothing but the elusive Compton cloud which 
reprocesses the incoming photons from a Keplerian disk and re-emits as high energy X-rays. However,
CENBOLs could be oscillating due to cooling processes (Molteni, Sponholz and Chakrabarti, 1996; Charabarti, Acharyya \& Molteni, 2004) 
or because of the topologies which do not permit standing shocks (Ryu, Molteni and 
Chakrabarti, 1997). The emissions from these oscillating CENBOLs naturally
explains the quasi-periodic oscillations observed in black hole candidates. There
are several artificial models of these QPOs in the literature, but ours are the solutions
of the most realistic governing equations till date. Furthermore, since our solutions show genericness of the physics
of oscillations, we predict that QPOs should be observed  even when the black hole 
is super-massive. For details regarding this and other simulations readers are
referred to Chakrabarti, Acharyya and Molteni (2004). 

\section{Results and Interpretations}

Numerical simulations have been carried out using the Smoothed Particle Hydrodynamics (SPH).
Both the upper and the lower halves of the equatorial planes have been used 
for injection of matter. The high Comptonization cooling by the disk with a
lower accretion rate is modeled by a power-law cooling from a flow of high accretion 
rate. This clever but appropriate simplification allowed us to resolve the mystery of 
QPOs in a most decisive manner. 

\begin{figure}[t]
\vskip 0.0cm
\centerline{\epsfxsize=5.0in\epsfbox{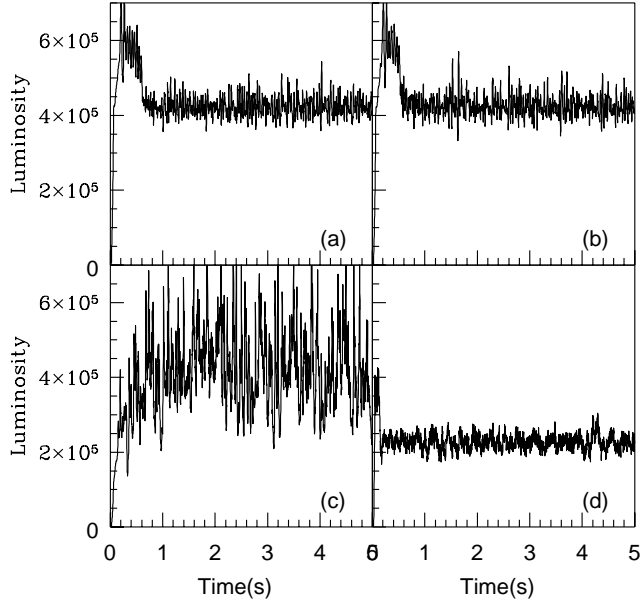}} 
\vskip -3.0cm
\caption{
Total luminosity of accretion flow (in arbitrary units) as a function of time (in seconds)
for injected matter density at the outer boundary (a) $\rho_{inj}=3.6\times 10^{-10}$gm/s, (b) $4.5
\times 10^{-10}$gm/s, (c) $4.5 \times 10^{-8}$gm/s and (d) $4.5 \times 10^{-7}$gm/s respectively.
The disk becomes cooler with the increase in accretion rate.
}
\end{figure}

Figures 1(a-d) show four light curves (luminosity vs. time) obtained by numerical simulations of low 
angular flows around black holes.  The accretion rates (assuming the cooling to be due to Comptonization)
were ${\dot m}= \frac{\dot M}{\dot M_{Edd}}= 0.05, 0.08, 0.39$ and $0.85$ respectively. As the 
accretion rates go up, the Compton $y$ parameter becomes high ($0.4$, $0.5$ and $11.8$)
and $43.4$. The cooling caused a drop in the post-shock pressure and the average 
location of the shock moved to $X_s= 15.8$, $16.3$, $24$ and $5$ respectively. 
After an initial transient state, the light curves showed typical fluctuations
as seen in $\chi$ state (Belloni et al. 2000) in GRS 1915+105. However, amplitude 
of fluctuation dropped when the ${\dot m}$ is close to $1$.

Figures 2(a-d) show the Power Density Spectrum of these light curves. We note that 
for low accretion rates, the quasi-periodic oscillation is very sharp (marked with arrows), and is located 
at the break frequency $\nu_b$. At the break frequency, the slope of the PDS abruptly changes
as in a observed data. The break frequency separates two types of flows: The flat power-law at lower frequency
($\nu<\nu_b$) is generated by the super-sonic pre-shock flow and the other part (for $\nu>\nu_b$) 
is generated by the subsonic CENBOL region located just outside the horizon. At high density, we observed
two QPO frequencies: one at $8.32$HZ and the other at $3.58$Hz respectively. The QPO frequencies depend on the
angular momentum and they basically scale inversely with the mass of the black hole.

\begin{figure}[t]
\vskip 0.cm  
\centerline{\epsfxsize=5.1in\epsfbox{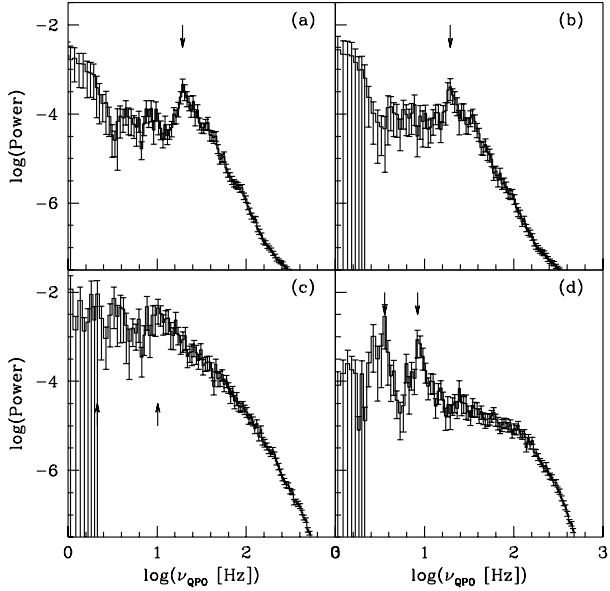}}
\vskip -3.0cm
\caption{Power density spectra for the light curves presented in Fig. 1. Arrows mark the 
frequencies at which the Quasi-Periodic Oscillations take place.  }
\end{figure}

\section{Concluding Remarks}

We showed that though black holes have no hard surfaces, the centrifugal barrier
in the flow can behave like one. The consequent CENBOL undergoes radial and vertical
oscillations. The X-rays emitted also undergo similar oscillations. The light curve
and PDS both have striking similarities with the observed light curve and PDS. We do not find any other
model in the literature which explains QPO so naturally. In GRS 1915+105, the 
behaviour is more complex that what is presented here. We believe that interaction of the
CENBOL, the base of the jet and the photons from the Keplerian disk will be important and small variations in accretion rates
of the Keplerian and sub-Keplerian flows will cause observed changes in light curve (Chakrabarti and Nandi, 2000). 
This is currently being investigated.

\section*{Acknowledgments}

SKC acknowledges a RESPOND project from Indian Space Research Organization (ISRO).

\end{document}